# On Shannon-Jaynes Entropy and Fisher Information

Vesselin I. Dimitrov[1]

*Idaho Accelerator Center, Idaho State University*
*1500 Alvin Ricken Dr.*
*Pocatello, ID 83201, USA*

**Abstract.** The fundamentals of the Maximum Entropy principle as a rule for assigning and updating probabilities are revisited. The Shannon-Jaynes relative entropy is vindicated as the optimal criterion for use with an updating rule. A constructive rule is justified which assigns the probabilities least sensitive to coarse-graining. The implications of these developments for interpreting physics laws as rules of inference upon incomplete information are briefly discussed.

## INTRODUCTION

### Motivation

Ever since E. Jaynes formulated the Maximum Entropy (MaxEnt) principle and used it to derive Statistical Thermodynamics [1], questions of the following type have developed significance and have been lingering around: To what extend other physics laws can be understood as rules of inference? How cheap can mechanics' first principles be [3]? How much of physics is just computation [2]? Are predictions of physics theories just those containing the least amount of information compatible with the descriptive framework used [4]? Is Quantum Mechanics the only consistent way to manipulate probability amplitudes [5]? And, last but not least, can *all of physics* be derived from information-theoretic principles [6]? In order to be able to even start thinking of answering these and similar questions, one needs a firm conceptual ground allowing unambiguous understanding of the MaxEnt principle itself. Unfortunately, we appear to still lack such an understanding [7]. Therefore, in the present work we set out to revise the fundamentals of the Maximum Entropy paradigm.

### The Paradigm and Its Shortcomings

*Constructive rule (MaxEnt)*

In its original scope [1] the MaxEnt principle is a constructive one - a device for assigning probabilities based on incomplete information. Probability assignments are thought to be adequate representation of one's state of knowledge about the system of interest, and the incomplete information is usually obtained by performing measurements on that system. The ultimate goal is to be able to predict the results of all possible measurements on the system from the outcomes of a finite, preferably small, number of measurements. This of course is not a well-posed problem in the usual sense, so the answer is sought in the form of probability distribution compatible with the measurements but otherwise as undedicated as possible. In order to put operational meaning into the vague notion of "as undedicated as possible" it is necessary to agree on a measure of information content (or the lack thereof) in a probability distribution. Once such a measure is selected, "most undedicated" translates into the "one with the least information content" compatible to the available information. For such a measure of the lack of information Jaynes chose to use the Shannon's entropy of a probability distribution $\{p_i, i = 1, \cdots, N\}$, namely $S[p] = -\sum_{i=1}^{N} p_i \ln p_i$, therefore we refer to this form, as well as to its generalizations in what follows, as the *Shannon-Jaynes (relative) entropy*. This choice is by no means unambiguous. Indeed, its usual justification is based on requiring the measure to obey a set of axioms and proving an existence and uniqueness theorem [8,9]. Such an axiomatic approach can be challenged on two counts - the plausibility of one or more of its axioms and the completeness of the proof that the proposed solution is unique. Unfortunately, the Shannon-Jaynes entropy characterization is susceptible to both these challenges. First, neither Shannon nor Khinchin made a very good point demonstrating that their distributivity axiom represents an absolutely necessary property of the measure, as discussed e.g. by Renyi [10]. Second, in proving the uniqueness of the measure Shannon apparently overlooks the circumstance that his argument holds for an arbitrary non-negative base of the logarithm function, including the possibility that different bins of the probability domain have different bases which may even depend on the bin's probability[2]. Thus a more general expression of the form

$$S[p] = -\sum_{i=1}^{N} p_i \log_{\alpha_i} p_i = -\sum_{i=1}^{N} p_i \ln \frac{p_i}{m_i},$$

with $m_i = p_i^{1-\ln \alpha_i}$ playing the role of apriori weights, results from the axioms. Introducing $\{m_i\}$ in place of the logarithm bases $\{\alpha_i\}$ merely facilitates the interpretation of these parameters as bin weights. That such weights are necessary becomes particularly obvious when one tries to pass to the limit of a continuous domain, and their arbitrariness forces a change on the interpretation of the MaxEnt prescription from a rule for assigning probabilities to a rule for updating probabilities.

---

[1] dimivess@isu.edu

*Updating rule (M.E.)*

Being interested mainly in the physics applications of the MaxEnt principle, in what follows we will consider probability distributions with continuous domains. In contrast to Shannon's original expression, the passage to continuum in the above formula is straightforward and results in a change-of-variable invariant measure

$$S[p] = -\int dx p(x) \ln \frac{p(x)}{m(x)}$$

---

[2] Khinchin, on the other hand, explicitly requires that the entropy is maximal for uniform probabilities, which effectively invokes the notorious Principle of Insufficient Reason and is open to all well-known objections to it [Ufink'95].

where $p(x)dx$ is the probability of $x \in [x, x+dx)$ and $m(x) > 0$. This is but the negative of the Kullback-Leibler (KL) distance between the probability distributions $p(x)$ and $m(x)$. In the new interpretation of the MaxEnt principle as an updating rule, which, following [11] I'll call "M.E.", one updates $m(x)$ to $p(x)$ by maximizing $S[p]$ (or, equivalently, minimizing the KL distance between $m$ and $p$) subject to constraints representing new information. In an attempt to circumvent the shortcomings of the Shannon's argument in the continuous case, Shore and Johnson [12] set out to derive an updating rule by imposing uniqueness and consistency requirements not on the form of the measure but directly on the outcome of the procedure. Although their system of axioms includes a requirement equivalent to simple additivity of the measure for independent subsystems, they surprisingly derived the Shannon-Jaynes form as the unique measure suitable for using in an updating rule. The surprise comes from the fact that, 20 years after Renyi published his paper, it should have been fairly well-known that simple additivity alone was too weak a condition to single out the Shannon-Jaynes relative entropy. Closer examination of Shore and Johnson's proof identifies a statement at the very end of their second Appendix where they argue that two functions, $F$ and $H$, are equivalent inasmuch as the one is a monotonic function of the other, so one can extremize either one, and they restrict their further considerations to $F$. What they apparently fail to recognize is that imposing the additivity requirement on $F$ is not the same as imposing it on $H$ and this is exactly what prevents them from obtaining the Renyi's family of relative entropies as further solution of their problem along with the Shannon-Jaynes relative entropy.

*Vàn's derivation*

Recently P.Vàn proposed a derivation of a constructive rule for the continuous case [13] where he allowed the information measure to depend on the first derivative of the probability density. This kind of dependence is a commonplace in variational calculus, but, apparently due to the discrete domain roots of the problem, has not been considered before in the MaxEnt context. Allowing the derivative to appear in the functional is also the most straightforward way to impose differentiability condition on the probability distribution function - a natural one in view of its physics applications. Regrettably, Vàn's derivation falls into the same trap as Shore and Johnson's in that being based on a simple additivity axiom it derives a measure in the form of a linear combination of the simple Shannon-Jaynes entropy and the Fisher information in $p$ associated with $x$. Although deficient, Van's derivation is the first to clearly indicate that Fisher information may play an important role in a rigorous MaxEnt approach.

There are further conceptual problems with the MaxEnt approach related to the way new information, obtained from measurements or otherwise, is introduced into the rule. They were first pointed at by Karbelkar [14], addressed by Ufink [7] and are discussed at length by Giffin and Caticha in these Proceedings. We will briefly mention some of those in the following sections.

To summarize, there are number of problems with the current MaxEnt paradigm which seem to justify a fresh, from scratch, approach to the derivation of updating and constructive rules. Such an approach is proposed in the subsequent two sections.

# CHARACTERIZATION OF THE UPDATING RULE

The rationale behind the M.E. updating rule is a very simple and sound one: Given that our state of knowledge regarding a system is encoded into a probability density $m(x)$ (designated "a prior") and new information is obtained in the form of an average value of system's observable, update $m(x)$ to $p(x)$ (called "posterior") such that the information distance between $m(x)$ and $p(x)$ is minimal, subject to the average value constraint. If this can be done in more than one equivalent ways the result should be the same. In other words, when updating probability distributions, one must be *conservative* and *consistent*. Clearly, all we need to implement the above desiderata is a proper definition of information distance. As discussed in the Introduction, when attempting to define information content in the usual way one necessarily ends up with information distance instead, so I'll try to axiomatically characterize the information content of a probability distribution in the hope to obtain an information distance measure.

## Basic Axioms

1. <u>(Weak) Locality</u>: The information content of a probability distribution $p(x)$ is an increasing function of a linear functional of $p(x)$ and $\nabla p(x)$ [3]:

$$S[p] = g(\int dx \tilde{f}(p,(\nabla p)^2))$$

   where $\tilde{f}(x,z)$ is a function to be determined;

2. <u>Expandability</u>: The information content is invariant upon expansion of the domain with zero probability assignments:

$$\int dx \tilde{f}(p,(\nabla p)^2) = \int dx p f(p,(\nabla p)^2) \text{ with } p f(p,(\nabla p)^2) \xrightarrow[p \to 0]{} 0$$

   where the unspecified $f(x,z)$ replaces $\tilde{f}(x,z)$;

3. <u>Additivity</u>: For independent probability distributions the information content of the joint distribution is the sum of the information contents of the individual distributions:

$$S[p_1(x)p_2(y)] = S[p_1(x)] + S[p_2(y)]$$

   This condition should restrict the particular form of the functions $g(x)$ and $f(x,z)$, hopefully unambiguously identifying them.

## Discussion of the Basic Axioms

The justification of the first axiom is given by Shore and Johnson [12]. We simply replace the locality requirement by a weak locality one in that dependence on the first derivatives is allowed. Including first derivatives in the linear functional restricts the original notion of a small variation of the probability density $p(x)$ from $\delta p(x): \max_x |\delta p(x)| = \varepsilon$ to $\delta p(x): \max_x |\delta p(x)| = \varepsilon, \max_x |\delta \nabla p(x)| = \varepsilon'$ thus enforcing differentiability upon $p(x)$. While the relevance of including differential criterion is not

---

[3] Vàn has shown that if one allows higher derivatives the form of the possible terms depends on the dimensionality of the configuration space, which is clearly undesirable. Therefore we restrict ourselves to first derivatives only.

immediately obvious in the discrete case, in the continuous case it cannot be apriori dismissed without further analysis.

The expandability axiom is a straightforward generalization of Khinchin's third axiom to the continuous case. In addition to its original justification we'll see later on that it is also a sufficient condition for the resulting updating rule to be consistent with the product rule of probability theory.

The additivity property is equivalent to what Shore and Johnson call "subset independence". While one can argue whether it is a necessary property of a general information content measure *per se*, it is absolutely essential for an updating rule based on such a measure in order to obtain consistent results when treating independent systems separately or combining them into a joint system.

## Consequences of the Basic Axioms

In view of the form of the expression $S[p] = g(\int dx p f(p, (\nabla p)^2))$ there are two possibilities for achieving additivity of $S[p]$: *i)* take $g(x+y) = g(x) + g(y)$ and require that $\int dx p f(p, (\nabla p)^2)$ is additive; or *ii)* take $g(xy) = g(x) + g(y)$ and require that $\int dx p f(p, (\nabla p)^2)$ factorizes. As it turns out, the first possibility is contained as a limiting case in the second one, so we only consider *ii)*. In this case $g(x)$ can be taken as $c \ln(x)$ where $c$ is an arbitrary constant. It is convenient to rewrite the so far unknown function $f(p, (\nabla p)^2)$ in the form $f(q, (\nabla q)^2)$ where $q(x) = \ln p(x)/m(x)$ and $m(x)$ is an arbitrary positive function. The advantage of this form is that for a joint distribution of independent variables both arguments of $f$ decompose to sums: $q(x,y) = q_1(x) + q_2(y)$, $(\nabla q(x,y))^2 = (\nabla_x q_1(x))^2 + (\nabla_y q_2(y))^2$. Thus it becomes easy to see that the factorization requirement

$$\int dx dy p_1 p_2 f(q_1 + q_2, (\nabla_x q_1)^2 + (\nabla_y q_1)^2) = \int dx p_1 f(q_1, (\nabla_x q_1)^2) \times \int dy p_2 f(q_2, (\nabla_y q_1)^2)$$

implies $f(q_1 + q_2, z_1 + z_2) = f(q_1, z_1) \times f(q_2, z_2) \Rightarrow f(q,z) = \exp(\alpha q + \beta z)$ with α and β arbitrary constants, and, finally

$$S[p] = c \ln \int dx p \left[ \left(\frac{p}{m}\right)^\alpha \exp[\beta(\nabla \ln \frac{p}{m})^2] \right]$$

This family, depending on three numerical parameters *c*, α, β and one arbitrary positive function *m(x)*, appears to be the most general form consistent with the basic axioms[4]. The presence of *m(x)* however indicates that the original intention of deriving a unique measure of information content, and thus a constructive rule, in this way, is doomed. In view of the fact that, for non-negative α, β and *c*, *S[p]* is non-negative and vanishes only for *p(x)=m(x)*, the alternative interpretation of *S[p,m]* as an information distance measure between the probability densities *p(x)* and *m(x)* suggests itself. Indeed, for particular choices of the parameters *S[p,m]* reduces to known forms of information distance: for β=0 and $c = -\alpha^{-1}$ $S[p,m] = -\alpha^{-1} \ln \int dx p (p/m)^\alpha$ is the Renyi distance, while for $\beta = b\alpha$, $c = a/\alpha$ and $\alpha \to 0$ $S[p,m] = a\int dx p \ln(p/m) + b\int dx p (\nabla \ln(p/m))^2$ is a Van-type distance; if further $b=0$ it reduces to the Shannon-Jaynes (or KL) distance, and if $a=0$ it turns into a kind of Fisher distance.

---

[4] A linear combination over the parameters with arbitrary weights will obviously also satisfy all requirements resulting from the axioms.

To summarize, the axiomatic approach doesn't seem capable of characterizing an information content measure but provides a characterization, albeit ambiguous one, of general purpose information distance measure. With such a characterization at hand, one can proceed to obtain an updating rule imposing goal-specific additional consistency requirements.

## Updating Rule: Narrowing Down the Choices

The use of the information distance measure derived above in a M.E. updating rule implies solving the variational problem $\delta\{S[p,m] + \lambda \int dx p(x)C(x)\} = 0$ where the variation is with respect to $p(x)$ and the Lagrange multiplier $\lambda$ is chosen such that $p(x)$ is consistent with the new information: $\int dx p(x)C(x) = d$. Instead of minimizing $S[p,m]$ one can equivalently minimize $\exp(c^{-1}S[p,m])$ so the general variational problem becomes

$$\delta\left\{\int dx p f(q, (\nabla q)^2) + \lambda \int dx p(x)C(x)\right\} = 0.$$

Let us now consider the case of two possibly dependent variables $x$ and $y$, a prior knowledge in the form of a joint probability distribution $m(x,y)$ and new information about $y$ to the effect that $y=y_0$. This is the same as constraining the marginal probability density of $y$ to a delta distribution $p(y) = \delta(y - y_0)$, which can be achieved by imposing infinitely many simple constraints, one for each $y$ value:

$$\int dy \lambda(y) \int dx p(x,y) = \delta(y - y_0).$$

Under the circumstances, any reasonable updating rule must produce a posterior of the form $p(x,y) = m(x|y)\delta(y-y_0) = m(x,y)\delta(y-y_0)/m(y)$ in compliance with probability theory's product formula for conditional probabilities. Working out the variational equation for this case

$$\left[1 + \alpha - 2\beta\Delta q - 2\beta(1+q\alpha)\frac{(\nabla q)^2}{q} - 4\beta^2 \nabla q \cdot \nabla\nabla q \cdot \nabla q - 2\beta\nabla \ln m \cdot \nabla q\right]f = \lambda(y)$$

it is easily seen that a solution with $q = p(x,y)/m(x,y) = q(y)$ is only possible, because of the last term in the brackets, for $\beta = 0$. Thus, requiring consistency with the probability theory's product rule eliminates the derivative terms from the information measure and leaves us with the family of Renyi measures. It is worthwhile noticing that, as mentioned in the Introduction, the fact that it is possible at all to choose the parameters in compliance with probability theory's product rule hinges essentially on the particular form of the functional required in Axiom 2 above.

## The Unique Updating Rule

Consider the following situation: There is a common knowledge about two variables $x$ and $y$ which is codified into a joint prior $m(x,y)$. A researcher $A$ acquires a "$g(x)$ meter" and uses it to measure the average value of $g(x)$. He then duly uses M.E. to update his probability distribution from $m(x,y)$ to $p_A(x,y)$. Another researcher, $B$, is fortunate enough to borrow a "*r(y)* meter" from a friend and uses it to measure the average value of *r(y)*. Then she similarly updates her prior to $p_B(x,y)$. Both researchers promptly publish the results of their measurements and thus become aware of each other's measured averages. *A* uses *B*'s result to update his probability distribution $p_A(x,y)$ to $p_{AB}(x,y)$. Analogously, *B* uses *A*'s result and updates her probability distribution $p_B(x,y)$ to $p_{BA}(x,y)$. Then there is a slacker *C* who doesn't care to experiment himself but carefully follows the literature and learns about *A* and *B*'s measurements. Using them simultaneously, he updates his prior $m(x,y)$ to $p_{(AB)}(x,y)$. For the sake of consistency, we require that $p_{AB}(x,y) = p_{BA}(x,y) = p_{(AB)}(x,y)$. Turning to the variational equation resulting from the choice $\beta = 0$ above one observes that the updating amounts to multiplying the prior with a certain function of the constraint:

$$p(x, y) = m(x, y) F_\alpha(\lambda C(x, y))$$

This being the case it is obvious that the first part of our consistency requirement is already satisfied - *A* and *B* will end up with identical probability distributions:

$$p(x, y) = m(x, y) F_\alpha(\lambda_1 g(x)) F_\alpha(\lambda_2 r(y))$$

As for the second part, it requires that the function $F_\alpha(z)$ has the property $F_\alpha(z_1 + z_2) = F_\alpha(z_1) F_\alpha(z_2)$ which is only the case for α=0 when it is an exponential. Thus, our three basic axioms plus the two consistency conditions uniquely single out the Shannon-Jaynes relative entropy as the one to use in M.E. updating. One has to be aware though that if in the above example the researchers measured different functions of the same variable, it would be in general impossible to obtain posteriors consistent between the three of them. This is precisely the problem pointed out by Karbelkar and discussed by Giffin and Caticha in these proceedings. M.E. with the Shannon-Jaynes relative entropy is the closest one can get to a consistent inference scheme, but it, too, falls short of being exactly that. Giffin and Caticha advocate for using M.E. with uninformative priors and keeping track of all previously used constraints. When new information becomes available, the corresponding constraint are to be combined with the old ones and the whole set employed to obtain a posterior. This obviously invalidates the fundamental tenet of the updating rule that prior knowledge should be encoded into a prior probability distribution, and thus renders the prior distributions obsolete.

## DERIVATION OF THE CONSTRUCTIVE RULE

In the preceding section we obtained a general information distance measure and used arguments derived from its intended use in a M.E. updating procedure to narrow down the choice of the β parameter value to β=0, thus excluding derivative terms. In fact, a much more general justification for this choice can be given by observing that the effective value of β depends on the scale on which the variable *x* is measured. Indeed, rescaling $x \to sx$ implies $\nabla \to s^{-1}\nabla$ and thus, the ratio *p/m* being invariant, $\beta \to s^{-2}\beta$. If one is to have a unique information distance measure, the only way to lift the ambiguity is to put β=0. This choice leaves us with a one-parameter family of "reasonable" information distance measures of the Renyi's type. Now we set out to use this family for the purpose of working out a rule for assigning probabilities based on incomplete information - the original Jaynes goal. The departure point is the following pair of axioms.

## The Axioms and the Constructive Rule

1. It is possible to meaningfully define a information content *I[p]* of a probability distribution *p(x)* so that $I[p] \in [0,\infty)$ and $|I[p_1] - I[p_2]| = L(S_\alpha[p_1, p_2])$ where *L()* is some convex linear combination over α;
2. The information content decreases upon coarse-graining: $I[p_\sigma] < I[p]$ where $p_\sigma(x)$ is $p(x)$ coarse-grained with a coarse-graining scale σ;

Consider now the amount of the decrease $0 < I[p] - I[p_\sigma] = L(S_\alpha[p, p_\sigma])$. Since the information content cannot go negative, the relation $I[p] - I[p_\sigma] = I[p]F(I[p])$ should hold with some function *F(z)* subject to the condition $zF(z) \xrightarrow[z \to 0]{} 0$. Inspecting both sides of the relation above it is easily seen that the probability density distribution for which the information content *I[p]* is minimal can be found among the probability density distributions for which $I[p] - I[p_\sigma]$, or equivalently, the information distance between the original and the coarse-grained probability density $S_\alpha[p, p_\sigma]$, is minimal. Thus, the following rule for assigning probabilities suggests itself: <u>When incomplete information is available, the probability distribution to be assigned is the one least sensitive to coarse-graining subject to all known constraints.</u>

## The Variational Equation

In working out the particularities of the constructive rule a great simplification results from utilizing the well-known fact that coarse-graining is statistically equivalent to adding noise. Indeed, the expression for the coarse-grained probability density $p_\sigma(x) = \int dz \, p(x + \sqrt{\sigma} z) f(z)$ where $f(z)$ is a non-negative normalized coarse-graining kernel and σ is the coarse-graining scale, can be interpreted as a marginalization $p_\sigma(x) = \int dz \, p_\sigma(x|z) f(z) = \int dz \, p_\sigma(x, z)$ from a joint probability density $p_\sigma(x, z)$ describing a noisy system. In the limit $\sigma = 0$ there is no coarse-graining and $p_{\sigma=0}(x, z) = p(x) f(z)$. We now consider the information distance between the original and the noisy probability distributions for infinitesimal coarse-graining (σ<<1): $S_\alpha[p_{\sigma=0}, p_\sigma] = \alpha^{-1} \ln \int dxdz \, p_{\sigma=0}(x, z) [p_{\sigma=0}(x, z) / p_\sigma(x, z)]^\alpha$.

Noticing that for isotropic coarse-graining kernel $\int dz \, z_i f(z) = 0$ and $\int dz \, z_i z_j f(z) = \chi \delta_{ij}$ where χ is a constant of the order of unity, we obtain by expanding in the powers of $\sigma^{1/2}$

$$S_\alpha[p_{\sigma=0}, p_\sigma] = \alpha^{-1} \int dxdz \, f(z) p(x) [p(x + \sqrt{\sigma} z) / p(x)]^{-\alpha} =$$
$$= \sigma \chi (\alpha + 1) \int dx \, p^{-1}(x) [\nabla p(x)]^2 + O(\sigma^2)$$

Remarkably, irregardless of the value of α[5], our probability assignment rule calls for a constrained minimization of the Fisher information

$$\int dx p^{-1}(x)[\nabla p(x)]^2 + \lambda \int dx p(x) C(x) \to \min$$

that would produce the *p(x)* least sensitive to coarse-graining subject to the known constraints. The corresponding Euler-Lagrange equation is easily obtained by setting $p(x) = \psi^2(x)$ and varying $\psi(x)$ without having to worry about the non-negativity of *p(x)*, to obtain

$$-\Delta \psi(x) + \lambda C(x) \psi(x) = 0$$

## DISCUSSION

Our revision of the MaxEnt fundamental principles vindicated the Shannon-Jaynes relative entropy form as the information distance measure which produces the closest thing to a consistent rule for updating probabilities. This does not come as a big surprise, given the documented enormous success of its numerous applications. The analysis, however, also indicates the impossibility of deriving a probability assignment rule from the usual requirements for consistency and uniqueness. New ideas are needed, and the strategy proposed above is not unlike the way the energy is treated in Field Theory: although we cannot define a particular finite form of an information content measure, the mere assumption that such a measure exists allows us to constructively address the question of locating its minimum. The derived constructive rule involves minimizing a particular instance of the Fisher information subject to data constraints. This result is especially interesting in view of our original motivation of establishing a relation between physics laws and information theory. While the role Fisher information plays in many variational principles of physics has been known and extensively discussed for a long time [6,16,17], no plausible explanation of why exactly the Fisher information should be used has been offered. The result of the previous section seems to provide such an explanation. It also can give insight into why physical systems appear to follow this or that particular law. For example, comparing the constructive rule with the Schrodinger equation of Quantum Mechanics $-\Delta \psi(x) + (2m/\hbar^2)[E - U(x)]\psi(x) = 0$, it appears that the ground state probability distribution of quantum systems is obtained by merely constraining the average kinetic energy, or equivalently the temperature, of the system to have a certain value as if quantum systems were in contact with a universal thermostat. The difference from the conventional statistical physics is that instead of the average kinetic energy value, the value of the Lagrange multiplier is prescribed to be proportional to the system's mass, namely $2m/\hbar^2$. Studying the why's and how's of this and similar connections could greatly advance our understanding of the origin of the physics laws and the inner workings of the universe around us.

---

[5] The statement hold even for linear combinations of different alphas, so the resulting rule is independent of whether one uses a Renyi distance, the symmetrized KL distance or even the Bernardo-Rueda divergence [15]